# Possible Extragalactic Sources of the Highest Energy Cosmic Rays[a]

Jörg P. Rachen

*Max-Planck-Institut für Radioastronomie*
*53010 Bonn, Germany*

## INTRODUCTION

The observation of cosmic ray events clearly above 100 EeV by the Fly's Eye[1] and AGASA[2] experiments on the one hand, and further evidence for an extragalactic origin of cosmic rays above about 3 EeV on the other hand (see below), create a dilemma for cosmic ray physics: because of the inevitable existence of the Greisen-Zatsepin-Kuzmin (GZK) cutoff at 60 EeV for protons and about 100 EeV for nuclei sources of such "super-GZK" cosmic rays have to be nearby, if not inside the Galaxy. Because the bending of particle paths in galactic and extragalactic magnetic fields over short distances is expected to be small, the sources should be observable somewhere close to the event directions, and it has been claimed that this is not the case.[3,4] However, it depends on the particles considered what "nearby" and "close" means, and it is not so easy to make a search for extragalactic sources complete in the galactic disk direction, where the Fly's Eye event was observed.

## WHY EXTRAGALACTIC?

Obviously, the problem we discuss may be constructed by the claim that cosmic rays above 3 EeV are of extragalactic origin. We therefore want to give some arguments that support this claim.

First, the statistically significant observation of a simultaneous change in spectrum and composition of cosmic rays at the "ankle" at about 3 EeV by the Fly's Eye[5] strongly suggests a change of components of different origin. The light and flat component shows a spectral slope of $-2.7$ *above* the ankle; but if we deduce its spectrum *below* the ankle from composition data, we find that here the slope is near to $-2.0$.[6] Even though this steepening of the light component is less significant, it is interesting to note that this is exactly what is expected for extragalactic cosmic rays due to Bethe-Heitler losses.

Second, in spite of the observation of the highest energy events the existence of a GZK "break" is not in contrast to the data: most experiments show a lack of particles above 60 EeV, compared with the expectation for the case of a continuing $E^{-2.7}$ spectrum.[2,5] Some experimental groups claim the existence of a "gap" between 60 and 200 EeV, because they did not observe events in this region. However, the Volcano Ranch, Haverah Park and Yakutsk arrays did observe in total 6 events slightly above 100 EeV.[7] In any case, the rather large energy errors allow a continuous distribution of the highest energy events as well, and, within the statistical flux errors, the "world data set" seems to be consistent with a spectral steepening at 60 EeV, which is expected for a universal source distribution with at least one local source to provide the highest energy events.[8]

[a]This work was developed in part at the Bartol Research Institute, University of Delaware, Newark DE. Work of the author at the Bartol Research Institute was supported by a DAAD Doktorandenstipendium HSP-II/AUFE.

Third, even if a gap exists (which would also suggest that the sub-GZK cosmic rays are extragalactic), the new component arising at 200 EeV is unlikely to be galactic, because all presently known acceleration models for compact objects fail to reach these energies by orders of magnitude, and an acceleration on galactic scales is unable to produce the extremely flat spectrum required for a rapid recovery.

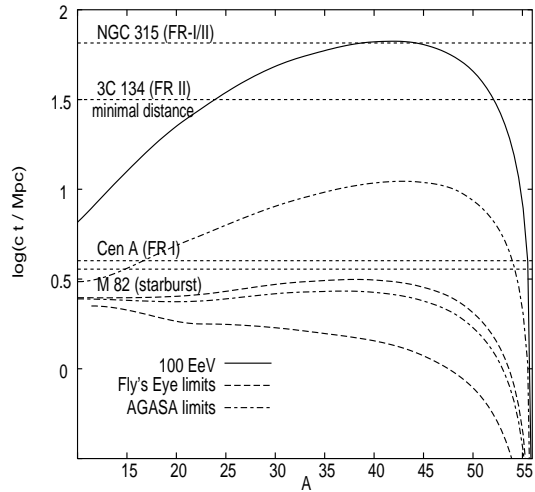

**FIGURE 1**: Range of iron nuclei, observed with mass $A$, for energies corresponding to the nominal GZK value for nuclei and the 1-$\sigma$ statistical limits of the highest energy events.

## RANGE AND DEFLECTION OF COSMIC RAY PARTICLES

At ultra high energies any particle may produce hadronic showers, which are more or less consistent with the data. Only neutrinos are expected to interact, if at all, much deeper in the atmosphere; however, particle physics at the interaction energy under discussion is far from being explored. We therefore are free to express the heuristic hypothesis of the existence of background immune particles (BIPs) with large hadronic cross section at extremely high energies. We now can distinguish the following possibilities:

*Protons*, with an average range of about 50 Mpc per decade of energy loss (above 300 EeV), and a 3-$\sigma$ upper limit allowed by fluctuations about three times larger. The maximum deflection is less than 10°.

*Heavy nuclei*, which allow for large deflection, but a smaller range at high energies following FIGURE 1, and only small fluctuations.

*Photons*, propagating on straight paths and limited by interactions with the radio background to an average range of about 20 Mpc at 300 EeV (3-$\sigma$ upper limit: 100 Mpc), strongly decreasing for smaller energies.

*BIPs*, which have an infinite range by definition. To be immune to photon interactions, they obviously have to be neutral and therefore point to their sources.

## COSMIC RAY SOURCES AND THEIR ENERGY LIMITS

Basically we may distinguish between two kinds of sources for highest energy cosmic rays: *accelerators* and *quantum processes.*

Accelerators are usually easy to observe by the radio emission of accelerated electrons. Radio luminosity may thus be used as a measure for acceleration efficiency, but it depends on the proton-to-electron ratio $k_p$. FR-II galaxies, the strongest extragalactic radio sources, can not only accelerate cosmic ray protons and nuclei up to a few 100 EeV in their hot spots, but also give the correct flux and spectral slope to explain the sub-GZK cosmic rays,[9] assuming a value $k_p$ consistent with the hadronic explanation of other observations.[10–13] Some of the weaker FR-I radio galaxies and compact jets in AGN may also be able to reach the energies,[14] but charged particles leaving the galaxy suffer adiabatic losses. Secondary neutral particles, however, can escape, thus FR-I galaxies may contribute energetic neutrons, and beamed jets in flat spectrum radio quasars (FSRQs) may supply neutrinos up to some 100 EeV.[15] Starburst galaxies, due to their high supernova rate, are expected to be sources of energetic heavy nuclei, which may reach highest energies.

For accelerating sources the cosmic ray flux can be estimated from observations. In contrast, the injection of energetic particles by decay of quantum-cosmological relics from the early universe, i.e. topological defects (TD), is much less observationally constrained.[4,16] These sources should appear as gamma pulses, observable above 100 EeV and maybe simultaneous in the TeV regime, but no systematic observations exist yet. Moreover, the energetics of the model has been questioned.[17]

## CORRELATION AND CONSEQUENCES

The assumptions about particles *and* their sources fall into two classes, depending on the continuation of the cosmic ray spectrum beyond the GZK break: if it just steepens, we can assume classical particles (protons and nuclei) originating in classical sources (radio or maybe starburst galaxies); if it cuts off and resumes at about 200 EeV, we probably have to assume a new particle component (photons or BIPs), arising from a new source population.

If the spectrum steepens we have the following choices: For the Fly's Eye event, there is the FR-II radio galaxy 3C 134 close to it, of which the redshift is unknown because no optical counterpart has been found due to galactic obscuration. From the size of the radio structure, we can estimate its distance between 30 and 300 Mpc, making it a possible proton source candidate. Close to the AGASA event we find the FR-I galaxies 3C 31 and NGC 315, which are inside the 3-$\sigma$ upper limit range of protons originating with >500 EeV. Considering heavy nuclei, we are less restricted in direction, and nearby starburst galaxies, as M 82 and NGC 253, can be discussed.

If there is a recovery of the spectrum, the new component may consist of photons from topological defects. However, the speculative BIP hypothesis opens a new possibility, based on an interesting observational coincidence: *both* the Fly's Eye and the AGASA event directions include a bright quasar in their error boxes; the FSRQ 3C 147 at $z \simeq 0.5$, and the radio weak PG 0117+213 at $z \simeq 1.5$, respectively. Since only a few 100 quasars with comparable brightness exist in the northern hemisphere, the probabilty to find this correlation by chance is less than 5%. While 3C 147 could be a source of extremely energetic neutrinos, this is unlikely for radio weak quasars.[15] However, both populations are connected by the expectation that these are *the only* objects where we have free view to the "surface" of a black hole, which is probably one of the most likely places for unexpected physics to happen.

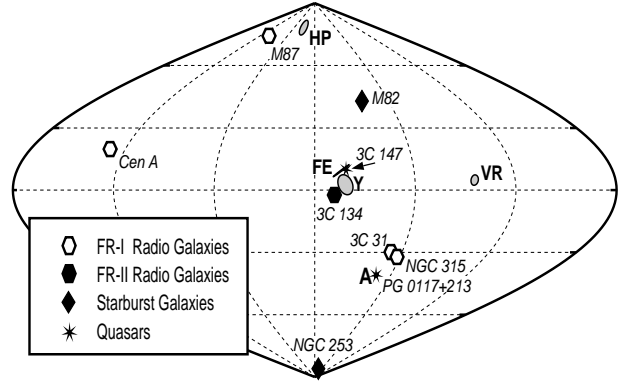

**FIGURE 2**: Distribution of the highest energy events of the experiments **A**GASA, **F**ly's **E**ye, **H**averah **P**ark, **V**olcano **R**anch and **Y**akutsk, and their possible sources discussed in the text, in galactic coordinates.

In conclusion, to constrain the speculations about the origin of the highest energy cosmic rays, we first need to get better event statistics above 100 EeV. If the high event energies and the high flux in this region can be confirmed, it would be interesting to check whether the correlation to quasar positions holds. Otherwise, if the events appear to be exceptional, they may be explained as the energetic tail of "normal" extragalactic cosmic rays.

## ACKNOWLEDGEMENTS

This work was developed in collaboration with Drs. T. Stanev and P.L. Biermann.[18]